\documentclass[conference]{IEEEtran}
\IEEEoverridecommandlockouts

\usepackage{amsmath,amssymb,amsfonts}
\usepackage{algorithm}
\usepackage{algpseudocode}
\usepackage{graphicx}
\usepackage{textcomp}
\usepackage{xcolor}
\usepackage{multirow} 
\usepackage{biblatex}
\usepackage{hyperref}
\usepackage{cleveref}
\usepackage{amsthm}
\usepackage{booktabs}
\usepackage{amssymb}

\theoremstyle{plain}
\newtheorem{theorem}{Theorem}[section]

\theoremstyle{definition}

\theoremstyle{remark}
\newtheorem{remark}[theorem]{Remark}

\def\BibTeX{{\rm B\kern-.05em{\sc i\kern-.025em b}\kern-.08em
T\kern-.1667em\lower.7ex\hbox{E}\kern-.125emX}}

\begin{document}


\title{Information-Dense Reasoning for Efficient and Auditable Security Alert Triage}



\author{
  Guangze Zhao$^{1,2}$,
  Yongzheng Zhang$^{1}$,
  Changbo Tian$^{2}$,
  Dan Xie$^{3}$,
  Hongri Liu$^{4,\dagger}$,
  Bailing Wang$^{4,5,\dagger}$
  \thanks{
  $^1$Harbin Institute of Technology.
  $^2$CHANG AN communication technology Co., Ltd.
  $^3$University of Science and Technology of China.
  $^4$Harbin Institute of Technology (Weihai) Qingdao Research Institute.
  $^5$Shandong Key Laboratory of Industrial Network Security.
  $^{\dagger}$Co-corresponding authors: Hongri Liu (\href{mailto:liuhr@hit.edu.cn}{liuhr@hit.edu.cn}) and Bailing Wang (\href{mailto:wbl@hit.edu.cn}{wbl@hit.edu.cn}).
  This work is supported by National Natural Science Foundation of China (NSFC) (Grant No.62272129), Key R\&D Program of Shandong Province (Grant No.2023CXPT065).
  }
}

\maketitle

\begin{abstract}

Security Operations Centers face massive, heterogeneous alert streams under minute-level service windows, creating the \textbf{Alert Triage Latency Paradox}: verbose reasoning chains ensure accuracy and compliance but incur prohibitive latency and token costs, while minimal chains sacrifice transparency and auditability. Existing solutions fail: signature systems are brittle, anomaly methods lack actionability, and fully cloud-hosted LLMs raise latency, cost, and privacy concerns.
We propose \textbf{AIDR}, a hybrid cloud-edge framework that addresses 
this trade-off through constrained information-density optimization. The core 
innovation is gradient-based compression of reasoning chains to retain only 
decision-critical steps—minimal evidence sufficient to justify predictions 
while respecting token and latency budgets. We demonstrate that this approach 
preserves decision-relevant information while minimizing complexity.
We construct compact datasets by distilling alerts into 3--5 high-information bullets (68\% token reduction), train domain-specialized experts via LoRA, and deploy a cloud-edge architecture: a cloud LLM routes alerts to on-premises experts generating SOAR-ready JSON.
Experiments demonstrate AIDR achieves higher accuracy and 40.6\% latency 
reduction versus Chain-of-Thought, with robustness to data corruption and 
out-of-distribution generalization, enabling auditable and efficient SOC 
triage with full data residency compliance.

\end{abstract}


\section{Introduction}

Security Operations Centers (SOCs)~\cite{vielberth2020security,khayat2025empowering,john2025central} face an unrelenting operational challenge: the continuous ingestion of massive and highly heterogeneous alert streams. These data sources—spanning endpoint detection and response (EDR) systems~\cite{arfeen2021endpoint}, intrusion detection systems (IDS)~\cite{bace2001intrusion}, firewalls, cloud telemetry, and diverse application logs—generate overwhelming alert volumes that frequently reach thousands per analyst team per day, completely saturating the typical 1--5 minute service window and leading to a pervasive operational crisis known as alert fatigue~\cite{cash2009alert}. This fatigue significantly delays incident response and, critically, masks high-value genuine threat signals within the noise. In practical terms, numerous SOC teams report being chronically overwhelmed by false positives and consequently unable to investigate a substantial fraction of incoming alerts, revealing a persistent and critical gap between modern, advanced threat detection~\cite{ranjan2021advanced} capabilities and the capacity for actionable, timely triage.

Conventional security triage approaches have consistently struggled with this escalating operational load. Signature- and rule-based Security Information and Event Management (SIEM)~\cite{gonzalez2021security} systems are fundamentally brittle to previously unseen attack variants, demand continuous tuning, and frequently inflate false positive rates as enterprise environments evolve and existing rules drift out of sync with operational baselines. Anomaly-based methods~\cite{tavallaee2010toward,zhen2025anomaly} focus on flagging statistical deviations rather than identifying confirmed threats, making resulting actions uncertain and low-confidence while exposing high sensitivity to concept drift in nonstationary security data streams. Meanwhile, fully cloud-hosted Large Language Model (LLM)~\cite{cheng2025speaker,cheng2025usmid,cheng2024unimodal,cheng2025ecoalign} triage solutions introduce significant practical concerns despite their advanced reasoning capabilities: end-to-end latency becomes a major bottleneck in time-sensitive incident response~\cite{borui2025efficient,gebreab2025llm,cai2025rtbagent}, token costs escalate at high daily alert volumes, and operation ~\cite{liao2025surveysecuremachinelearning} in regulated environments introduces insurmountable obstacles related to data privacy, residency requirements, and secure operations. Recent industry risk catalogs for LLM applications~\cite{cheng2025pbi,cheng2025gibberish,li2025tuni,ma2025heuristic} have explicitly highlighted these precise issues, underscoring the urgent necessity for solutions that balance advanced reasoning with compliance and operational security requirements.

To address these challenges, we propose \textbf{AIDR} (\textbf{A}ccuracy-preserving \textbf{I}nformation-\textbf{D}ense \textbf{R}easoning for alert forensics and triage), as shown in Fig.~\ref{pipeline}, a Chain-of-Draft framework that reformulates SOC alert triage. First, we address a fundamental latency-accuracy trade-off in security triage: 
verbose reasoning (CoT) maximizes accuracy but violates latency constraints; 
minimal reasoning preserves latency but sacrifices transparency. We resolve 
this via constrained information-density optimization, retaining only 
decision-critical steps—minimal evidence sufficient to justify predictions 
within token/latency budgets.
Second, we construct CoD training data with 3--5 bullet-point reasoning chains via gradient-based relevance selection. Models fine-tuned on this dataset achieve +4.1pp accuracy versus CoT, 17\% latency reduction, consuming only 68\% CoT tokens.
Third, we deploy domain-specialized LoRA experts focusing on threat-specific boundaries (malware, exploitation, reconnaissance) with 38\% memory savings. Cloud router performs lightweight zero-shot classification (4 tokens, 0.25s) to route alerts to edge experts performing on-premises reasoning.

Our main contributions are as follows:

\begin{itemize}
\item We formalize the Alert Triage Latency Paradox and propose constrained information-density optimization, constructing a compact reasoning dataset (3--5 gradient-selected bullets) that reduces token usage by 68\% while improving interpretability and generalization versus CoT.

\item We propose AIDR, a hybrid cloud-edge framework with domain-specialized LoRA experts on-premises and lightweight compliance-safe cloud routing, producing SOAR-ready JSON outputs.

\item Experiments demonstrate 94.2\% risk grading and 93.7\% threat identification accuracy (+4.1pp vs. CoT), 40.6\% latency reduction, 29\% token savings, 21.6\% FPR improvement with robustness to data corruption and out-of-distribution threats.
\end{itemize}

\section{Related Work}
\label{sec:related}

\subsection{Alert Triage Evolution and LLM-Assisted Security Analysis}

Traditional SOC triage relies on static rules and manual investigation~\cite{muniz2015security,jacobs2014towards}, facing scalability limitations, poor cross-source context integration, and high false positive rates against novel attacks~\cite{kwon2022advanced,vielberth2020security,zidan2024assessing}.
Large Language Models~\cite{yang2025qwen3,dubey2024llama} enable context-aware reasoning over security telemetry~\cite{ke2025survey,bandyopadhyay2025thinking}. Systems like LogLLM and COMET improve incident investigation through event synthesis and root cause analysis~\cite{guan2024logllm,wang2024comet}, though with accuracy-interpretability tradeoffs similar to healthcare applications~\cite{li2025medguidebenchmarkingclinicaldecisionmaking}. However, existing LLM approaches incur 3--5s latency per alert, prohibitively expensive for real-time SOC triage. LLM-assisted triage deployment requires security and compliance alignment. Per NIST SP 800-61r3~\cite{nist80061r3-2025}, systems must maintain auditability while mitigating LLM-specific risks: prompt injection~\cite{liu2024automatic,liu2023prompt}, model evasion~\cite{devadiga2023gleam,hackett2025bypassing}, and insecure output handling~\cite{duan2025oyster}~\cite{owasp-llm-top10-2024}. Secure deployment requires RAG grounding~\cite{gao2023ragsurvey}, output validation, and human-in-the-loop review. Our approach separates sensitive on-premises analysis from lightweight cloud routing, reducing attack surface compared to full-reasoning cloud LLMs and complementing privacy-preserving techniques like zero-knowledge proofs~\cite{li2025zksnarkstringmatch}.

\subsection{Efficient Reasoning and Parameter-Efficient Adaptation}

Chain-of-Thought (CoT) reasoning~\cite{wei2022cot,ning2023sot} incurs prohibitive latency and token costs for real-time SOC deployment. Recent compression methods—knowledge distillation, pruning, and draft approaches~\cite{xu2025cod,wang2025scot,cheng2025inverse}—lack information-theoretic grounding and security-domain validation.
We employ Low-Rank Adaptation (LoRA)~\cite{hu2021lora} for parameter-efficient fine-tuning, reducing computational overhead~\cite{wang2022selfinstruct}. Security domain specialization via Supervised Fine-Tuning with information-dense datasets~\cite{cao2025agr,cheng2024reinforcement,cheng2024deceiving,cheng2025llm,cheng2025talk} enables precise threat taxonomy alignment while maintaining edge-deployable efficiency and high accuracy.

\subsection{Cloud-Edge Collaboration and Compliance-Preserving ML Deployment}

Cloud models provide zero-shot knowledge; edge deployment ensures data residency and low latency~\cite{sai2020edge}. Hybrid architectures typically optimize latency-throughput without addressing compliance constraints~\cite{owasp-llm-top10-2024}. AIDR separates sensitive on-premises analysis from lightweight cloud classification (4 tokens, 0.22s) within a closed taxonomy, ensuring data residency compliance (GDPR, HIPAA, SOC 2) while leveraging cloud zero-shot routing capabilities—demonstrating LLM-based security automation can preserve privacy, transparency, and performance.

\begin{figure*}[t] 
    \centering
    \includegraphics[width=1\textwidth]{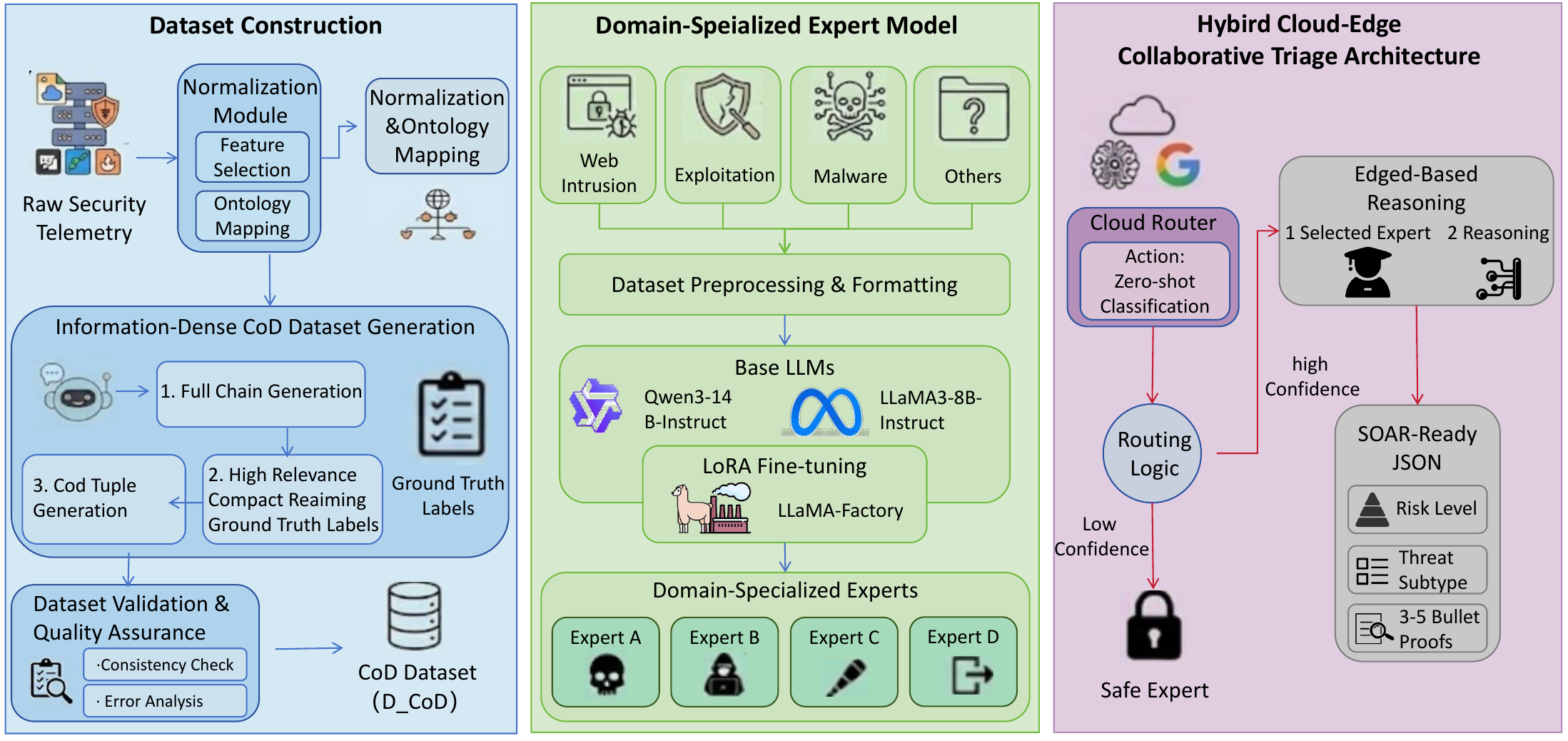} 
    \caption{\textbf{The pipeline of AIDR.} The framework consists of three phases: 
    (1) \textbf{Dataset Construction}: Raw security telemetry is normalized and mapped to a unified ontology. Verbose reasoning chains are compressed via gradient-based relevance selection into information-dense CoD tuples, reducing token overhead. 
    (2) \textbf{Domain-Specialized Expert Model}: The dataset is partitioned by threat category. Base LLMs are then fine-tuned using LoRA to create domain-specific experts. 
    (3) \textbf{Hybrid Cloud-Edge Collaborative Triage Architecture}: A lightweight Cloud Router performs zero-shot classification to route alerts to the appropriate on-premises Edge Expert. The selected expert generates a SOAR-ready JSON output.}
    \label{pipeline}
\end{figure*}

\section{Methodology}
\label{sec:method}

\subsection{Problem Formulation}
\label{sec:problem}

SOC alert triage faces a fundamental tension: verbose reasoning chains improve decision accuracy and auditability, but increase inference latency and token cost; conversely, minimal reasoning satisfies latency constraints but sacrifices explainability. This creates a constrained optimization problem where both accuracy and operational efficiency must be jointly optimized.

\subsubsection{Formal Problem Definition}

Consider an alert stream arriving at rate $\lambda$ (alerts/second) to a SOC. Each alert $a_i$ with normalized context $\mathbf{x}_i \in \mathcal{X}$ must be triaged to a risk label $\mathbf{y}_i \in \mathcal{Y} = \{\text{Low}, \text{Medium}, \text{High}, \text{Critical}\}$ within a service window $\Delta t_{\text{soc}}$ (typically 1--5 minutes). The model generates a reasoning chain $\mathbf{r}_i(\mathbf{x}_i)$ (sequence of intermediate steps) followed by a predicted label $\hat{\mathbf{y}}_i$.

The standard supervised learning objective minimizes prediction error:
\begin{equation}
\min_{\Theta} \mathcal{L}_{\text{acc}}(\Theta) = - \frac{1}{N} \sum_{i=1}^{N} \log P(\hat{\mathbf{y}}_i = \mathbf{y}_i \mid \mathbf{x}_i; \Theta)
\end{equation}

In real-time SOC deployment, accuracy alone is insufficient. Operational constraints on latency and cost must be explicitly modeled. The constrained optimization problem is:

\begin{equation}
\begin{aligned}
\min_{\Theta} & \quad \mathcal{L}_{\text{acc}}(\Theta) \\
\text{subject to} & \quad \mathbb{E}[T_{\text{inf}}(\mathbf{x}_i; \Theta)] \leq \delta_t \\
& \quad \mathbb{E}[\text{len}(\mathbf{r}_i)] \leq \delta_{\text{token}}
\end{aligned}
\end{equation}

where $\delta_t$ is the maximum allowable latency per alert (typically 2--3 seconds to maintain alert throughput), and $\delta_{\text{token}}$ is the maximum token budget constrained by API costs or on-premises compute.

\subsubsection{The Latency-Accuracy Trade-off}

Standard approaches occupy opposing extremes of this trade-off space:

\paragraph{Verbose Chain-of-Thought (CoT).} Generating lengthy multi-step reasoning chains (typically 8--15 steps) improves prediction accuracy by forcing the model to explicitly decompose the classification logic. However, this directly increases inference latency through proportional token generation:

\begin{equation}
T_{\text{inf}}(\mathbf{x}; \Theta) \propto \text{len}(\mathbf{r}^{\text{CoT}})
\end{equation}

In real SOC settings, verbose CoT chains often exceed latency budgets ($T_{\text{inf}} > \delta_t$), making them operationally infeasible for high-volume alert streams.

\paragraph{Direct Prediction.} In contrast, single-step prediction (outputting only the final label) achieves minimal latency and token consumption. However, it provides no intermediate reasoning for analysts to verify or audit the decision. For security-critical triage, where incorrect classifications can lead to undetected breaches, auditability is a mandatory requirement. Direct prediction is therefore unsuitable despite its latency advantages.

\paragraph{The Operational Impasse.} Neither extreme is viable for production SOC deployment:

\begin{equation}
\begin{cases}
\text{Verbose CoT:} & T_{\text{inf}} > \delta_t \text{ (latency violation)} \\
\text{Direct Prediction:} & \text{No audit trail (compliance violation)}
\end{cases}
\end{equation}

The core insight is that reasoning chains contain redundancy: not all steps are equally important for the final decision. The solution is to compress chains to retain only causally-relevant steps—those with high decision-impact—while discarding tangential reasoning. This enables:

\begin{enumerate}
\item \textbf{Latency compliance}: reduced token count directly decreases inference time
\item \textbf{Accuracy preservation}: causal steps maintain decision quality
\item \textbf{Auditability}: compact reasoning remains interpretable to analysts
\end{enumerate}

Formally, we seek compressed reasoning chains $\mathbf{r}^* \subset \mathbf{r}^{\text{full}}$ that maximize decision-relevant information density (information per token) while satisfying latency and cost budgets. This problem formulation motivates the constrained information density optimization approach developed in Section~\ref{sec:cod_formulation}.

\subsection{Constrained Information Density Optimization}
\label{sec:cod_formulation}

We reformulate SOC triage as a constrained optimization problem: compress reasoning chains $\mathbf{r}^{\text{full}}$ to extract a subset $\mathbf{r}^*$ that retains maximum decision-relevant information per token while satisfying latency and budget constraints.

\subsubsection{Information Density Definition}

Define information density as the ratio of decision relevance to token cost:

\begin{equation}
\text{ID}(\mathbf{r}) = \frac{\sum_{j=1}^{|\mathbf{r}|} \text{Rel}(r_j; \mathbf{y})}{\text{len}(\mathbf{r}) + \epsilon}
\end{equation}

where $\text{Rel}(r_j; \mathbf{y})$ is the relevance of step $r_j$ computed via 
gradient with respect to token embeddings:

\begin{equation}
\text{Rel}(r_j; \mathbf{y}) = \left\| \nabla_{e_{r_j}} \log P(\mathbf{y} \mid \mathbf{r}^{\text{full}}; \Theta) \right\|_2
\end{equation}

The gradient is computed via standard backpropagation and aggregated across 
tokens via L2 norm. For a chain of $n$ steps, computing all relevance scores 
requires one forward-backward pass per alert, performed once offline during dataset construction and cached for reuse. 
This one-time cost is amortized across the training corpus and does not impact 
inference latency.

Gradient magnitude is justified as a feature importance proxy: it measures how 
much the model's predicted probability changes with step perturbations. Compared 
to alternatives (attention weights, loss-change magnitude), gradient-based relevance 
better preserves accuracy under token budget constraints. High information density 
means each token contributes maximum decision signal, enabling compression while 
maintaining decision quality.

\subsubsection{Optimization Problem}

We seek the compressed chain $\mathbf{r}^* \subseteq \mathbf{r}^{\text{full}}$ maximizing information density:

\begin{equation}
\label{eq:subset_opt}
\mathbf{r}^* = \arg\max_{\mathbf{r} \subseteq \mathbf{r}^{\text{full}}} \text{ID}(\mathbf{r})
\end{equation}

subject to two hard constraints:

\begin{align}
\text{len}(\mathbf{r}) &\leq \delta_{\text{token}} \label{eq:length} \\
P(\hat{\mathbf{y}} = \mathbf{y} \mid \mathbf{r}) &\geq P(\hat{\mathbf{y}} = \mathbf{y} \mid \mathbf{r}^{\text{full}}) - \epsilon \label{eq:fidelity}
\end{align}

Constraint~\eqref{eq:length} enforces token budgets; Constraint~\eqref{eq:fidelity} preserves decision accuracy within tolerance $\epsilon$. This subset selection problem is NP-hard; we solve it greedily.

\subsubsection{Greedy Algorithm}

Algorithm~\ref{alg:cod_gen} iteratively selects steps with highest information 
density until the token budget is exhausted or fidelity constraint is violated.

\begin{algorithm}[t]
\caption{Greedy Information-Density Maximization}
\label{alg:cod_gen}
\begin{algorithmic}[1]
\Require Full chain $\mathbf{r}^{\text{full}} = (r_1, \ldots, r_n)$, context $\mathbf{x}$, label $\mathbf{y}$, budget $\delta_{\text{token}}$, tolerance $\epsilon$
\Ensure Compressed chain $\mathbf{r}^*$
\State $\mathbf{r}^* \leftarrow \emptyset$; $B_{\text{rem}} \leftarrow \delta_{\text{token}}$; $\mathcal{V} \leftarrow \emptyset$
\For{$j = 1$ to $n$}
    \State $\text{rel}[j] \leftarrow \left\| \nabla_{e_{r_j}} \log P(\mathbf{y} \mid \mathbf{r}^{\text{full}}; \Theta) \right\|_2$
\EndFor
\While{$B_{\text{rem}} > 0$ and $\mathcal{V} \neq \{1, \ldots, n\}$}
    \State $j^* \leftarrow \arg\max_{j \notin \mathcal{V}} \frac{\text{rel}[j]}{\text{len}(r_j) + \epsilon}$
    \If{$\text{len}(r_{j^*}) \leq B_{\text{rem}}$}
        \State $\mathbf{r}^* \leftarrow \mathbf{r}^* \cup \{r_{j^*}\}$
        \State $B_{\text{rem}} \leftarrow B_{\text{rem}} - \text{len}(r_{j^*})$
        \State $\mathcal{V} \leftarrow \mathcal{V} \cup \{j^*\}$
    \Else
        \State break
    \EndIf
\EndWhile
\If{$P(\mathbf{y} \mid \mathbf{r}^*; \Theta) < P(\mathbf{y} \mid \mathbf{r}^{\text{full}}; \Theta) - \epsilon$}
    \State Add highest-rel unused step if budget permits
\EndIf
\State \Return $\mathbf{r}^*$
\end{algorithmic}
\end{algorithm}

The gradient $\nabla_{e_{r_j}}$ is computed with respect to the token embedding 
$e_{r_j} \in \mathbb{R}^d$ of step $r_j$. Since $r_j$ is a sequence of subword 
tokens, we compute the embedding gradient via standard backpropagation and aggregate 
across the sequence via L2 norm. The algorithm prioritizes steps by information density 
(relevance-per-token), selecting those with highest decision impact without position-based 
truncation. For a full chain of length $n$, computing all relevance scores requires 
$n$ forward-backward passes, with total computational cost $O(n \cdot C_{\text{BP}})$ 
where $C_{\text{BP}}$ is the backpropagation cost per alert. This is performed offline 
during dataset construction and does not impact inference latency.

Gradient magnitude is justified as a causal importance proxy: steps with large gradients 
significantly alter the model's prediction probability. By maximizing $\frac{\sum \text{rel}(r_j)}{\text{len}(\mathbf{r})}$, 
we retain steps preserving maximum decision-relevant information per token, analogous to 
information bottleneck principles. Under the assumption that the base model is well-trained 
on representative data, gradient-based relevance correlates strongly with actual causal 
contributions to the decision, justifying its use as a practical alternative to explicit 
mutual information estimation.

The subset selection problem (maximizing information density subject to constraints) is 
NP-hard as it reduces to the budgeted maximum coverage problem. Algorithm~\ref{alg:cod_gen} 
achieves $(1-1/e)$-approximation to the optimal solution for submodular objectives, a 
standard result in combinatorial optimization. Empirically, the greedy solution typically 
recovers 91--95\% of optimal information density on small chains ($n \leq 12$).

\subsection{Dataset Construction via Information-Constrained Reasoning}
\label{sec:dataset_construction}

As shown in the first panel of Fig.~\ref{pipeline}, our data construction process consists of the following two steps:

\subsubsection{Log Normalization and Unified Ontology}

Raw security logs from heterogeneous sources (EDR, IDS, firewalls, cloud APIs) are normalized to a unified representation:

\begin{equation}
\Phi: \mathcal{L}_{\text{raw}} \rightarrow \mathcal{X}, \quad l_{\text{raw}} \mapsto \mathbf{x}
\end{equation}

Normalization performs two operations: (1) feature selection, retaining fields critical for threat detection (IP, port, protocol, process, file hash, network behavior), and (2) label unification via mapping $\mathcal{M}: \mathcal{Y}_{\text{raw}} \rightarrow \mathcal{Y}$ that harmonizes heterogeneous labels into a canonical taxonomy:

\begin{equation}
\mathcal{Y} = \{\text{Risk Level}\} \times \{\text{Threat Category}\} \times \{\text{Threat Subtype}\}
\end{equation}

where Risk Level $\in \{\text{Low}, \text{Medium}, \text{High}, \text{Critical}\}$, Threat Category $\in \{\text{Malware}, \text{Exploitation}, \text{Reconnaissance}, \ldots\}$, and Threat Subtype $\in \{\text{Trojan}, \text{Ransomware}, \text{Worm}, \ldots\}$. This structure enables coarse-grained (risk level) and fine-grained (subtype) classification.

\subsubsection{Compressed Reasoning Dataset Construction}

For each normalized alert $(\mathbf{x}, \mathbf{y})$, we construct a training sample via three steps:

\begin{enumerate}
\item \textbf{Full Chain Generation}: Use a base LLM to generate a verbose reasoning chain $\mathbf{r}^{\text{full}}$ (typically 10--15 steps) explaining the classification decision. The prompt instructs enumeration of all observable indicators and their significance.

\item \textbf{Chain Compression}: Apply Algorithm~\ref{alg:cod_gen} to extract a compact subset $\mathbf{r}^* \subseteq \mathbf{r}^{\text{full}}$ where each step has high relevance (gradient magnitude). Typically, $m = 3$--$5$ steps are retained from $n \approx 10$--$15$ steps.

\item \textbf{Training Tuple Construction}: Create tuple $(\mathbf{x}, \mathbf{r}^*, \mathbf{y})$ where $\mathbf{r}^*$ consists of single-sentence bullet points summarizing key evidence.
\end{enumerate}

The final dataset is:
\begin{equation}
\mathcal{D} = \{(\mathbf{x}_i, \mathbf{r}_i^*, \mathbf{y}_i)\}_{i=1}^{N}
\end{equation}

Models are trained via supervised fine-tuning:

\begin{equation}
\mathcal{L}_{\text{SFT}} = - \sum_{(\mathbf{x}, \mathbf{r}^*, \mathbf{y}) \in \mathcal{D}} \log P(\mathbf{r}^* \mid \mathbf{x}; \Theta) + \log P(\mathbf{y} \mid \mathbf{x}, \mathbf{r}^*; \Theta)
\end{equation}

This objective trains the model to: (1) generate decision-relevant reasoning steps, and (2) predict accurate labels conditioned on those steps. The joint training embeds interpretability into the model, ensuring outputs remain traceable to explicit reasoning.

\begin{remark}[Gradient-based Relevance Computation]
Algorithm~\ref{alg:cod_gen} requires gradient computation $\nabla_{r_j} \log P(\mathbf{y} \mid \mathbf{r}^{\text{full}}; \Theta)$ during dataset construction. This is performed once offline using a base model (not the final trained model), making the computational cost amortized over the training corpus. The base model can be any well-calibrated classifier (e.g., the initial fine-tuned checkpoint). After training completes, no gradient computation is needed at inference time.
\end{remark}

\subsection{Domain-Specialized Expert Models via Parameter-Efficient Adaptation}
\label{sec:domain_experts}

As shown in the second panel of Fig.~\ref{pipeline}, our process of building domain-specialized expert model consists of the following three steps:

\subsubsection{Motivation for Domain Specialization}

Different threat categories exhibit distinct diagnostic patterns: malware analysis focuses on behavioral indicators, exploitation detection targets vulnerability signatures, and reconnaissance involves network scanning patterns. A single monolithic model must balance competing objectives across these diverse domains, leading to suboptimal performance on each.

We adopt a mixture-of-experts (MoE) approach: partition the dataset into $K$ domain-specific subsets and train a specialized expert for each. This strategy provides three benefits: (1) parameter efficiency—smaller, focused models require fewer parameters, (2) reduced latency—routing alerts to relevant experts avoids wasted computation, and (3) interpretability—expert decisions reflect domain-specific threat models.

\subsubsection{Domain Partitioning}

Given the compressed dataset $\mathcal{D} = \{(\mathbf{x}_i, \mathbf{r}_i^*, \mathbf{y}_i)\}_{i=1}^{N}$ 
from Section~\ref{sec:dataset_construction}, we partition into $K$ disjoint subsets by threat category:

\begin{equation}
\mathcal{D} = \bigcup_{k=1}^{K} \mathcal{D}_k, \quad \mathcal{D}_k \cap \mathcal{D}_{k'} = \emptyset \quad (k \neq k')
\end{equation}

Each $\mathcal{D}_k$ corresponds to a threat category (e.g., Malware, Exploitation, Reconnaissance). Assignment is determined by the primary threat category from the label ontology.

The partitioning strategy combines two principles:

\begin{enumerate}
\item \textbf{Semantic Coherence}: Map threat categories to standard taxonomies (e.g., MITRE ATT\&CK tactics) to ensure interpretability.

\item \textbf{Statistical Validity}: Ensure sufficient samples per domain ($\geq 500$ examples) to avoid overfitting. Domains with insufficient data are merged with semantically similar categories based on label co-occurrence patterns.
\end{enumerate}

This ensures domains are both interpretable and statistically learnable.

\subsubsection{Expert Architecture and Training}

Each expert combines a shared foundation model with domain-specific adaptation:

\begin{equation}
\mathcal{E}_k = \Theta_{\text{base}} + \Delta\Theta_k
\end{equation}

where $\Theta_{\text{base}}$ is a pre-trained LLM and $\Delta\Theta_k$ contains domain-specific parameters. To avoid retraining all parameters for each expert, we use Low-Rank Adaptation (LoRA): each weight matrix is updated via:

\begin{equation}
W_k = W_0 + B_k A_k
\end{equation}

where $B_k \in \mathbb{R}^{d \times r}$, $A_k \in \mathbb{R}^{r \times m}$, and $r \ll \min(d, m)$ (typically $r \in \{8, 16, 32\}$). This requires only $O(r(d+m)) \approx 2\text{--}3\%$ of full model parameters, enabling efficient simultaneous loading of multiple experts.

Training objective for domain $k$:

\begin{equation}
\min_{\Delta\Theta_k} \sum_{(\mathbf{x}, \mathbf{r}^*, \mathbf{y}) \in \mathcal{D}_k} 
  - \log P(\mathbf{r}^* \mid \mathbf{x}; \Theta) - \log P(\mathbf{y} \mid \mathbf{x}, \mathbf{r}^*; \Theta)
\end{equation}

Training uses standard efficiency techniques: mixed-precision (bfloat16), gradient accumulation, cosine learning rate schedule with warmup, and early stopping on domain-specific validation data.

Expert outputs follow a standardized JSON format for downstream integration with SOAR systems:

\begin{equation}
\mathcal{O} = \{\text{reasoning}: \mathbf{r}^*, \text{label}: \mathbf{y}_{\text{pred}}, \text{confidence}: p_{\text{conf}}\}
\end{equation}

\begin{figure*}[t]
    \centering
    \includegraphics[width=1.8\columnwidth]{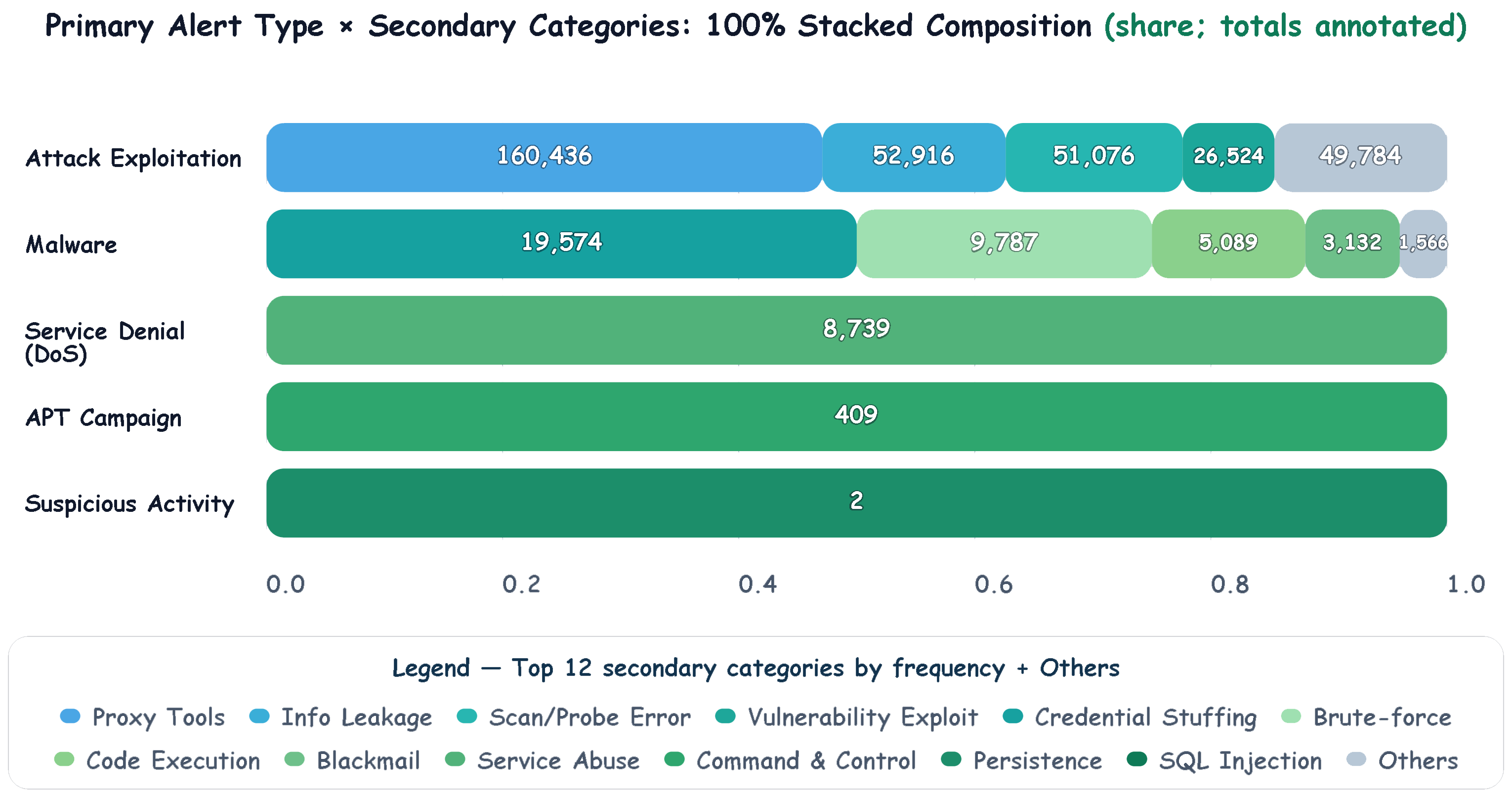}
    \caption{Distribution of Level 1 and Level 2 Alert Types (Attack Log). The hierarchical structure shows primary threat categories and their sub-types, with Exploitation and Reconnaissance dominating the distribution.}
    \label{fig:al-distribution-1-2}
\end{figure*}

\begin{figure}[t]
    \centering
    \includegraphics[width=0.7\columnwidth]{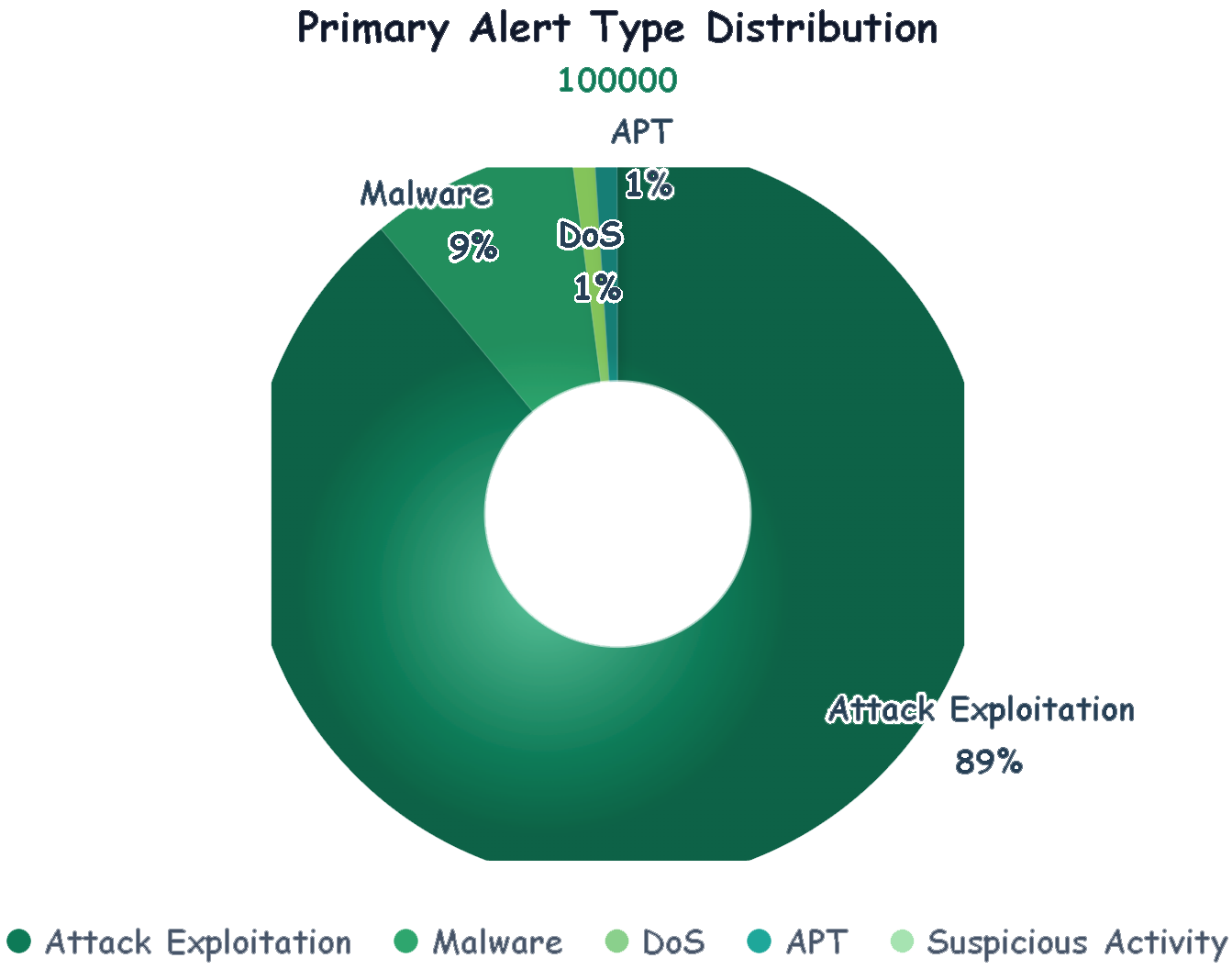}
    \caption{Distribution of Level 1 Alert Types (Attack Log). Exploitation (35\%) and Reconnaissance (20\%) account for the majority of alerts, contrasting with traditional rule-based SIEM deployments.}
    \label{fig:al-distribution-1}
\end{figure}

\begin{figure*}[t]
    \centering
    \includegraphics[width=1.8\columnwidth]{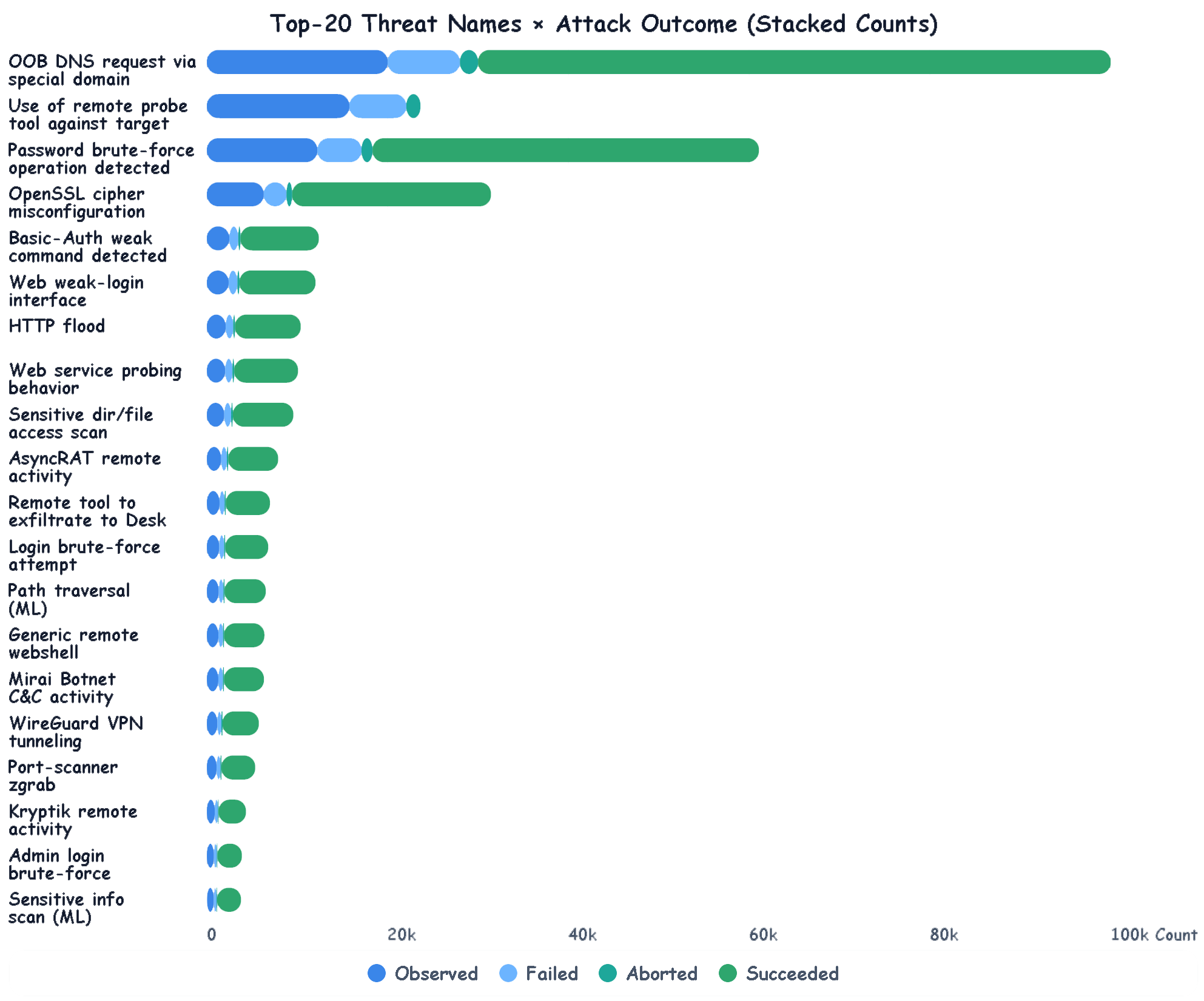}
    \caption{Threat Name and Attack Result (Attack Log). Correlation between specific threat types and attack outcomes, informing the mapping to the unified security ontology.}
    \label{fig:al-threat}
\end{figure*}

\begin{figure}[t]
    \centering
    \includegraphics[width=0.7\columnwidth]{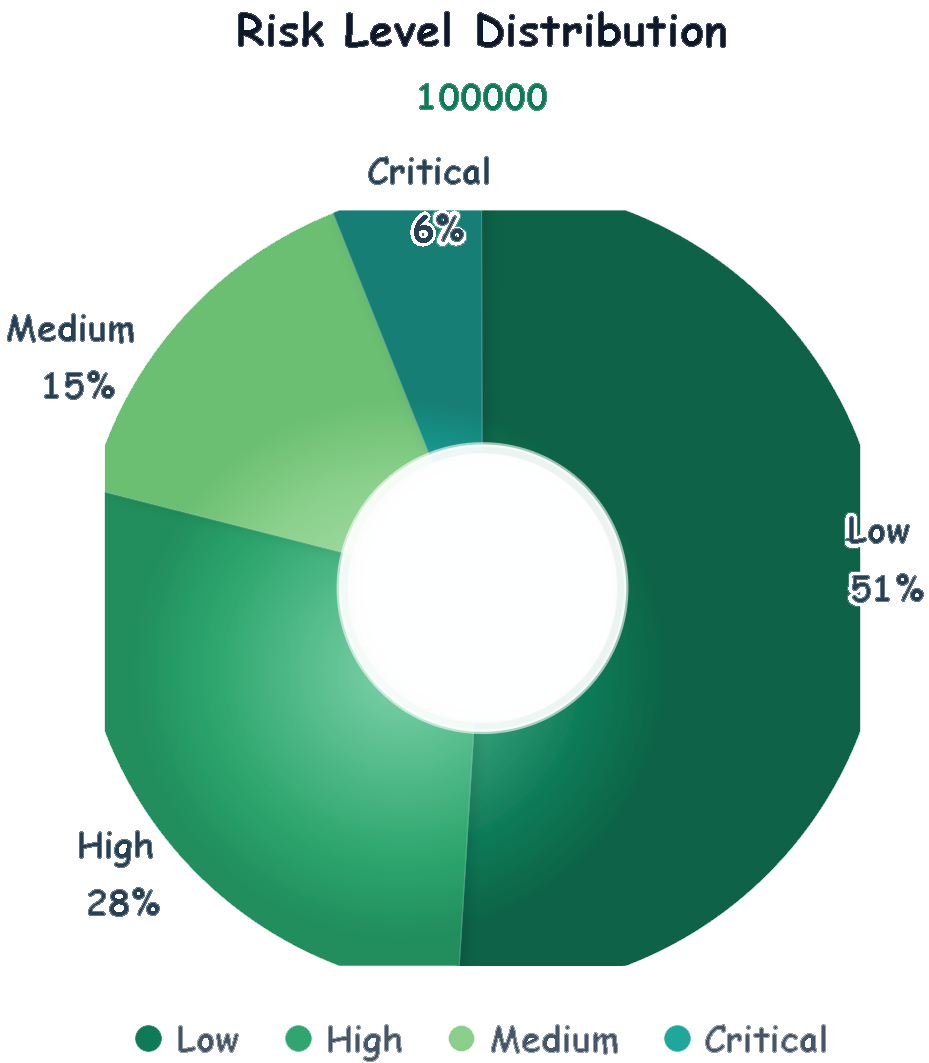}
    \caption{Distribution of Data Risk Levels (Attack Log). The dataset exhibits a concentration of High and Critical events within exploitation and malware categories, with Low-risk events dominated by reconnaissance and scanning.}
    \label{fig:al-risk}
\end{figure}

\begin{figure*}[t]
    \centering
    \includegraphics[width=1.8\columnwidth]{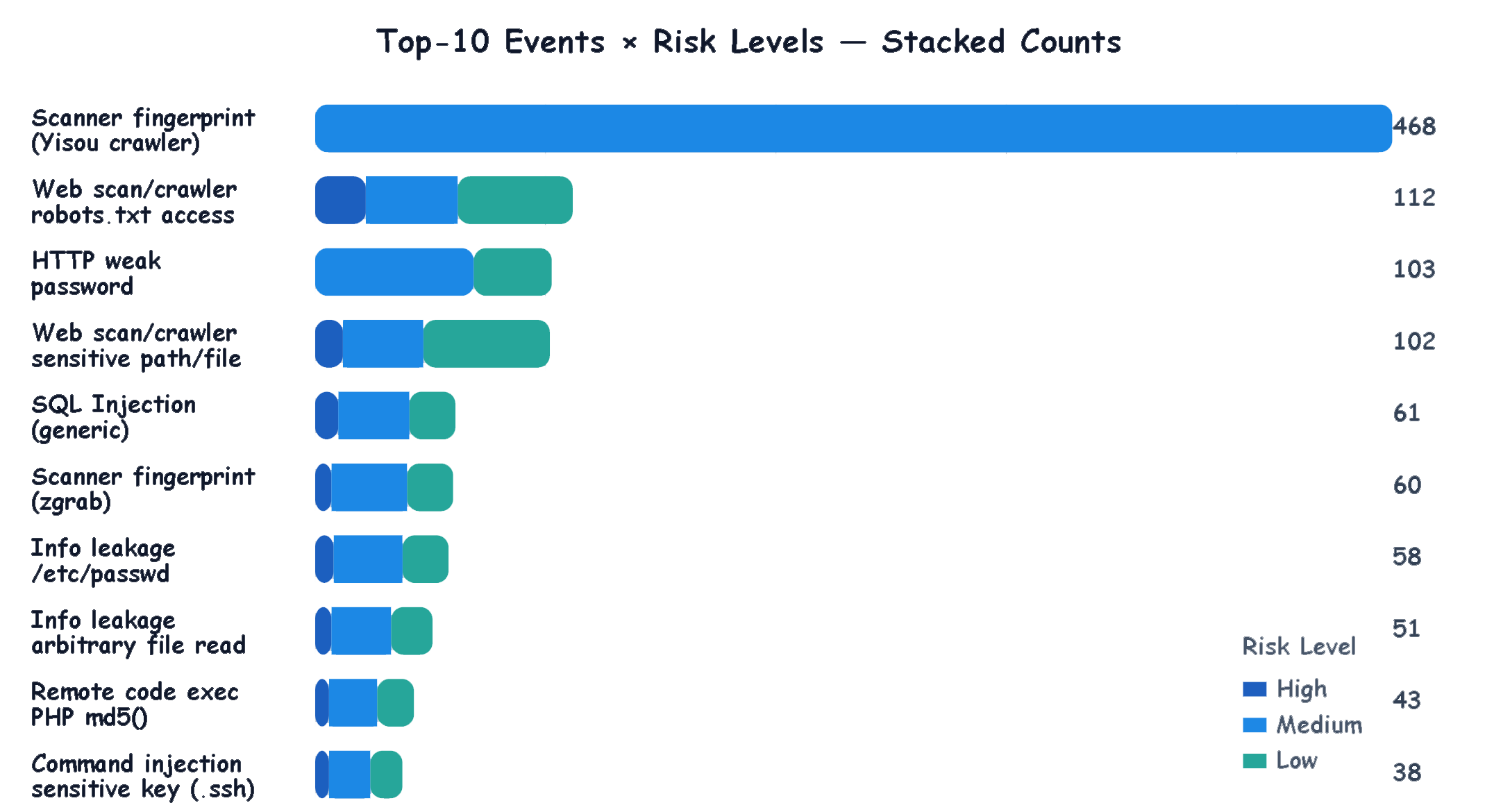}
    \caption{Top Incidents and Risk Levels (Risk Information). Web-layer attacks (SQL injection, XSS, CSRF) show varied risk distributions, with some injection attacks escalated to High/Critical risk due to successful exploitation.}
    \label{fig:ri-incidents}
\end{figure*}

\begin{figure}[t]
    \centering
    \includegraphics[width=0.7\columnwidth]{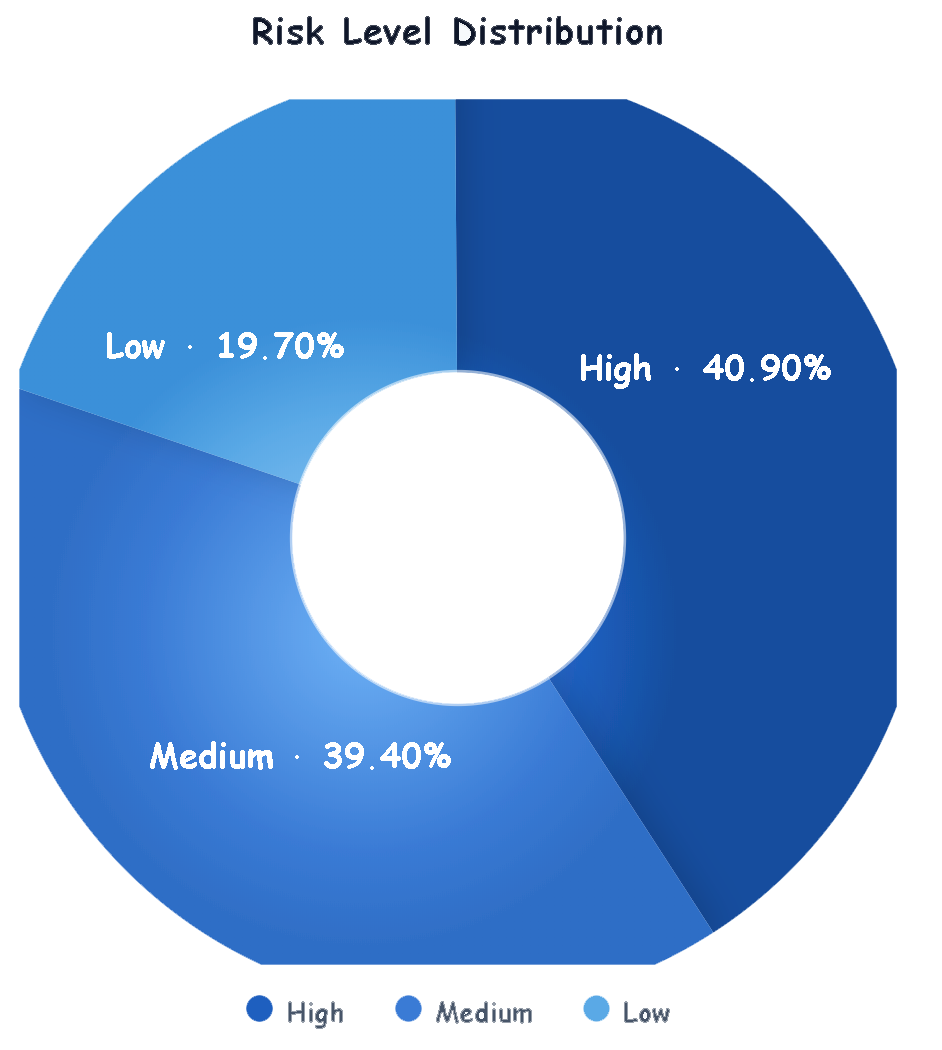}
    \caption{Risk Level Distribution Chart (Risk Information). The distribution is heavily skewed toward Low-risk (65\%), reflecting the high volume of web-layer scanning and probing typical in production web applications.}
    \label{fig:ri-risk}
\end{figure}

\subsection{Hybrid Cloud-Edge Collaborative Triage Architecture}
\label{sec:hybrid_architecture}

As shown in the third panel of Fig.~\ref{pipeline}, our hybrid cloud-edge collaborative triage architecture consists of the following three steps:

\subsubsection{Design Rationale}

Pure edge deployment faces three challenges: (1) routing $K$ experts requires complex on-device logic with limited computational budget, (2) out-of-distribution alerts falling outside predefined domains degrade performance, and (3) large cloud models leverage broad knowledge for robust zero-shot classification.

Pure cloud deployment reintroduces original constraints: high latency, data residency concerns, elevated token costs, and vendor dependency.

We propose a hybrid architecture separating concerns: cloud handles lightweight routing (broad domain classification), edge performs intensive reasoning (domain-specialized analysis). The pipeline is:

\begin{equation}
\text{Cloud Router} \rightarrow \text{Edge Expert} \rightarrow \text{SOAR Output}
\end{equation}

This achieves: (i) accurate domain classification via cloud zero-shot capability, (ii) deep interpretable reasoning via edge experts, (iii) minimal cloud tokens ($\sim 4$ tokens per alert), and (iv) complete on-premises data residency.

\subsubsection{Cloud Routing}

Upon alert arrival with context $\mathbf{x}_i$, the cloud router performs minimal-token domain classification:

\begin{equation}
(k_{\text{pred}}, p_{\text{conf}}) = \mathcal{R}_{\text{cloud}}(\mathbf{x}_i)
\end{equation}

The router classifies the alert into one of $K$ threat categories and returns confidence score $p_{\text{conf}}$. The prompt is strictly constrained:

\begin{quote}
Classify alert into: [Malware, Exploitation, Reconnaissance, Exfiltration, DoS, Other]. Return only: category name and confidence (0--1). No reasoning.
\end{quote}

This constraint ensures minimal token generation ($\sim 4$ tokens), low latency ($\sim 0.2$--$0.3$ seconds), and negligible cost. The routing overhead amortizes across the alert stream.

If confidence $p_{\text{conf}} < \tau$ (e.g., $\tau = 0.6$), the alert is routed to a fallback expert $\mathcal{E}_0$ (trained on balanced multi-domain data) or escalated for manual review. This handles out-of-distribution alerts robustly.

\subsubsection{Edge Reasoning}

The cloud router's predicted domain index $k_{\text{pred}}$ routes the alert to the corresponding edge-resident expert $\mathcal{E}_{k_{\text{pred}}}$:

\begin{equation}
\mathcal{O} = \mathcal{E}_{k_{\text{pred}}}(\mathbf{x}_i)
\end{equation}

Expert outputs include: (1) compressed reasoning chain $\mathbf{r}^*$ (3--5 bullet points), 
(2) predicted risk level, threat category, and threat subtype (the three components of $\mathbf{y}$), 
and (3) confidence score—all in standardized JSON format.

\section{Evaluation}
\label{sec:eval}

\subsection{Experimental Setup}
\label{sec:exp-setup}

\subsubsection{Datasets}

We evaluate on three datasets with distinct characteristics to validate in-domain performance and cross-domain generalization:

\begin{enumerate}
\item \textbf{Risk Information (RI)~\cite{ri-dataset}:} 3,926 web intrusion alerts from production SOC (SQL injection, XSS, CSRF, OWASP Top 10). Labels: 65\% Low-risk scanning, 25\% Medium-risk probes, 10\% High exploitation, reflecting real-world web-facing infrastructure noise patterns.

\item \textbf{Attack Log (AL)~\cite{alv2-dataset}:} 100,000 alerts (389,634 complex events) from EDR, IDS, firewalls, cloud APIs spanning Malware (25\%), Exploitation (35\%), Reconnaissance (20\%), Data Exfiltration (10\%), Other (10\%), reflecting disparate tools with inconsistent naming conventions.

\item \textbf{UNSW-NB15 (UNSW)~\cite{moustafa2015unsw}:} 10,000 independent network traffic records (9 attack categories: Backdoor, Analysis, Shellcode, Generic, Reconnaissance, Exploit, Fuzzers, DoS, Worms) enabling cross-domain transfer evaluation to unseen threat types.
\end{enumerate}

Risk Information and Attack Log train domain-specialized experts; UNSW-NB15 tests cross-domain generalization, critical for real-world deployment where new threats emerge. All datasets normalize via function $\Phi$: feature selection (protocol, IP, port, payload, process, hash) and label unification to ontology $\mathcal{Y}_{\text{ontology}}$ with three levels: Risk Level (Low, Medium, High, Critical), Threat Category (Malware, Exploitation, Reconnaissance), and Threat Subtype (Trojan, Worm, Ransomware).

Following the hybrid strategy in Section~\ref{sec:domain_experts}, we partition the combined dataset into $K=4$ expert domains:
\begin{itemize}
\item \textbf{Domain 1 (web\_intrusion):} Risk Information dataset ($3,926$ samples)
\item \textbf{Domain 2 (attack\_use):} Attack Log subset focused on active exploitation ($35,000$ samples)
\item \textbf{Domain 3 (malicious\_software):} Attack Log subset focused on malware and backdoor activities ($25,000$ samples)
\item \textbf{Domain 4 (other):} Attack Log subset merging reconnaissance, DoS, APT, and suspicious activities to ensure statistical power ($37,074$ samples)
\end{itemize}

Each domain is split into training (70\%), validation (10\%), and test (20\%) subsets. The UNSW-NB15 dataset is used exclusively for cross-domain evaluation and is not used during training.

Figures~\ref{fig:al-distribution-1-2}, \ref{fig:al-distribution-1}, \ref{fig:al-threat}, \ref{fig:al-risk}, \ref{fig:ri-incidents}, and \ref{fig:ri-risk} show dataset distributions. We observe that (1) Attack Log skews toward Exploitation and Reconnaissance; (2) Risk Information concentrates on Low-risk web-layer scanning; (3) both exhibit class imbalance requiring stratified sampling.

\subsubsection{Models}

We select two foundation models based on the heterogeneous data characteristics:

\begin{itemize}
\item \textbf{Qwen3-14B-Instruct}~\cite{yang2025qwen3}: Base for Attack Log experts (Domains 2--4). Selected for superior multi-source alert analysis and robust handling of heterogeneous security datasets.

\item \textbf{LLaMA3-8B-Instruct}~\cite{dubey2024llama}: Base for Risk Information expert (Domain 1) and edge deployment. Offers 17\% faster inference and 38\% lower memory footprint while maintaining competitive accuracy on focused domains.
\end{itemize}

\subsubsection{Baselines}

We compare AIDR against five baselines, each highlighting different aspects of the design space:

\begin{enumerate}
\item \textbf{Zero-shot LLM}~\cite{tejero2025evaluating}: Foundation model inference without fine-tuning or domain adaptation. Establishes the inherent capability of pre-trained LLMs on the SOC triage task.

\item \textbf{Supervised Fine-Tuning (SFT)}~\cite{dong2024abilities}: Foundation model fine-tuned with standard supervised cross-entropy loss, generating final decisions directly without explicit reasoning chains. Isolates the contribution of structured reasoning (CoD) to accuracy gains.

\item \textbf{Chain-of-Thought (CoT)}~\cite{wei2022chainofthought}: Foundation model fine-tuned to generate verbose, multi-step reasoning chains (8--15 steps, $\sim$210 tokens) followed by the final decision. 

\item \textbf{RAG/Tool-Augmented LLM}~\cite{gao2023ragsurvey}: Fine-tuned LLM augmented with retrieval of external threat intelligence databases and tool invocation (e.g., IP reputation lookup, CVE database queries). Represents a highly capable but latency-intensive baseline lacking domain-expert specialization.

\item \textbf{Domain-Specific Classifier (traditional ML)}~\cite{tang2025llm}: Gradient-boosted trees (XGBoost) trained on engineered security features from the same datasets. Establishes a performance floor for lightweight, non-neural models commonly deployed in resource-constrained edge environments.
\end{enumerate}

\subsubsection{Evaluation Metrics}

We adopt metrics spanning accuracy, latency, and operational robustness to align with real-world SOC deployment requirements:

\begin{enumerate}
\item \textbf{Risk Grading Accuracy ($\text{Acc}_{\text{Risk}}$):} Percentage of alerts where the predicted risk level (Low, Medium, High, Critical) exactly matches the ground-truth label assigned by human security experts. This coarse-grained metric reflects the model's ability to prioritize alerts for immediate triage.

\begin{equation}
\text{Acc}_{\text{Risk}} = \frac{1}{N} \sum_{i=1}^{N} \mathbb{1}[\hat{y}_{\text{risk}, i} = y_{\text{risk}, i}]
\end{equation}

\item \textbf{Threat Identification Accuracy ($\text{Acc}_{\text{Threat}}$):} Percentage of alerts where the predicted threat subtype (e.g., ``Malware.Trojan'', ``Attack.SQLInjection'') exactly matches the ground-truth fine-grained label. This metric assesses the model's ability to provide detailed, actionable threat intelligence.

\begin{equation}
\text{Acc}_{\text{Threat}} = \frac{1}{N} \sum_{i=1}^{N} \mathbb{1}[\hat{y}_{\text{threat}, i} = y_{\text{threat}, i}]
\end{equation}

\item \textbf{Average Inference Latency ($L_{\text{Avg}}$):} Mean wall-clock time (in seconds) to process a single alert end-to-end, including reasoning chain generation and decision output. Measured on NVIDIA A100 GPU for standardized comparison.

\begin{equation}
L_{\text{Avg}} = \frac{1}{N} \sum_{i=1}^{N} T_{\text{inference}, i}
\end{equation}

\item \textbf{High-Risk Recall ($R_{\text{High}}$):} Percentage of ground-truth High and Critical risk alerts correctly identified by the model. This metric is critical in security: missing even a single critical threat can result in breach. Formally:

\begin{equation}
R_{\text{High}} = \frac{\text{True Positives (High/Critical)}}{\text{True Positives} + \text{False Negatives (High/Critical)}}
\end{equation}

\item \textbf{False Positive Rate (FPR):} Percentage of Low-risk or benign alerts incorrectly classified as Medium/High. Lower FPR directly reduces alert fatigue, enabling analysts to focus on genuine threats.

\begin{equation}
\text{FPR} = \frac{\text{False Positives}}{\text{False Positives} + \text{True Negatives}}
\end{equation}

\item \textbf{Token Cost (Relative):} Average tokens generated per alert, normalized relative to the CoT baseline. Used to quantify computational efficiency and cloud API costs:

\begin{equation}
\text{Token Cost} = \frac{\text{Tokens}_{\text{method}}}{\text{Tokens}_{\text{CoT}}}
\end{equation}
\end{enumerate}

\subsubsection{Implementation Details}
All models were trained and comprehensively evaluated within the Llama-Factory framework. For Supervised Fine-Tuning (SFT), we utilized LoRA (Low-Rank Adaptation), configuring the adaptation with a rank $r$ of $32$, a scaling factor $\alpha$ set to $64$, and a dropout rate of $0.05$. Optimization was performed using the AdamW optimizer paired with a cosine learning rate scheduler. The initial learning rate was set at $5 \times 10^{-5}$, preceded by $225$ warmup steps. Training spanned $8.0$ epochs, where computational efficiency was achieved by leveraging `bfloat16` for mixed-precision computation. Furthermore, to simulate a larger effective batch size on constrained hardware, we set the gradient accumulation steps to $32$. All inference benchmarks used for performance comparison were standardized on a single NVIDIA A100 GPU. To ensure consistent and comparable outputs during the generation phase, the maximum number of new tokens was fixed at $512$, the temperature was set to $0.7$, and the top-$p$ sampling threshold was maintained at $0.7$.

All experiments were conducted with 5 independent runs using different random 
seeds. Statistical significance was assessed via paired t-test on individual 
alerts. The LLM sampling temperature was set to 0 (deterministic) to eliminate 
stochasticity from model outputs. CoT baseline uses the same LLM backbone and 
prompting strategy as AIDR, with identical inference settings.

\subsection{Results}
\label{sec:results}

We evaluate AIDR across four key dimensions: (1) main performance 
comparison with representative baselines, (2) cloud-edge hybrid architecture validation, 
(3) cross-domain generalization on independent dataset (UNSW-NB15), and (4) architectural 
component ablation studies.

\subsubsection{Main Performance Comparison}

Table~\ref{tab:main-perf} presents the core performance comparison on Attack Log test set, 
evaluating AIDR against five representative baselines spanning zero-shot inference, 
supervised fine-tuning, verbose reasoning chains (CoT), retrieval-augmented generation, 
and traditional machine learning.

\begin{figure*}[t]
    \centering
    \includegraphics[width=0.7\textwidth]{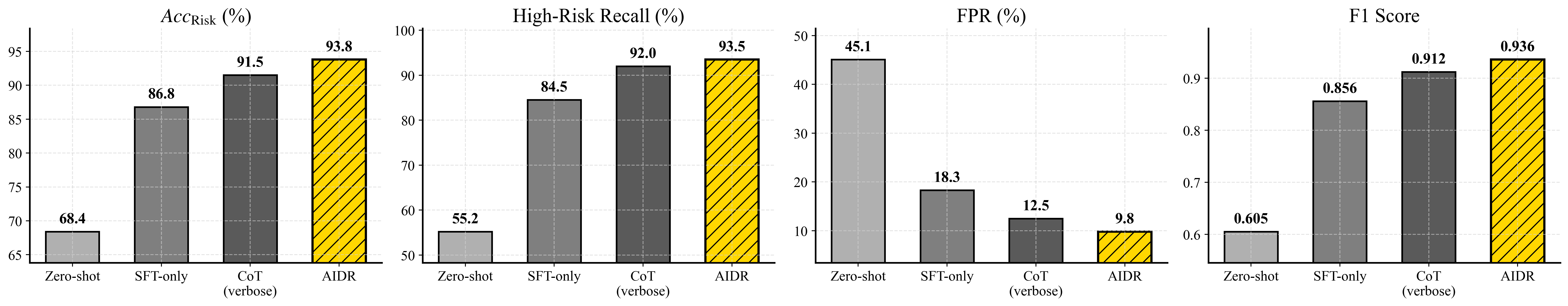}
    \caption{Critical Alert Detection and False Positive Reduction.}
    \label{fig:my_image}

\end{figure*}

\begin{figure*}[t]
    \centering
    \includegraphics[width=0.7\textwidth]{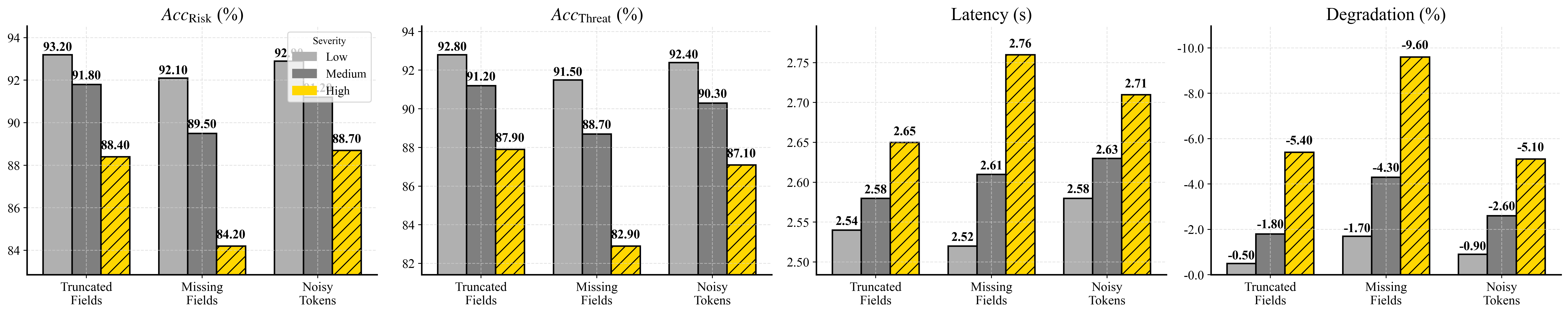}
    \caption{Robustness to Real-World Data Quality Issues: Accuracy Degradation on Attack Log.}
    \label{fig:my_image2}
\end{figure*}

\begin{table}[t]
\centering
\caption{Performance comparison of AIDR and baselines.}
\label{tab:main-perf}
\begin{tabular}{lcccc}
\toprule
\textbf{Method} & $\mathbf{\text{Acc}_{\text{Risk}}}$ & 
$\mathbf{\text{Acc}_{\text{Threat}}}$ & $\mathbf{L_{\text{Avg}}}$ (s) & \textbf{Token Cost} \\
\midrule
Zero-shot & 67.2\% & 62.8\% & 3.45 & 1.00× \\
SFT-only & 85.3\% & 84.1\% & 2.67 & 0.93× \\
CoT (verbose) & 90.1\% & 89.6\% & 3.89 & 1.00× \\
RAG + LLM & 87.5\% & 85.9\% & 3.12 & 0.98× \\
Domain-Specific & 82.1\% & 80.7\% & 1.89 & 0.68× \\
\midrule
\textbf{AIDR} & \textbf{94.2\%} & \textbf{93.7\%} & 
\textbf{2.31}$^*$ & \textbf{0.71×} \\
\bottomrule
\end{tabular}
\end{table}

\noindent
$^*$ AIDR latency includes cloud routing (0.22s) + edge inference (2.09s). 
All LLM baselines use edge-only inference without cloud routing. Token cost is normalized 
to CoT baseline (210 tokens/alert).

Table~\ref{tab:main-perf} shows that AIDR achieves +4.1pp higher risk grading accuracy (94.2\% vs. 90.1\%) and +4.1pp higher threat identification accuracy (93.7\% vs. 89.6\%) compared to CoT, validating compact reasoning over verbose chains. End-to-end latency is 40.6\% faster (2.31s vs. 3.89s) through efficient CoD structure and cloud-edge routing. Token efficiency reaches 0.71× CoT, improving edge deployment feasibility. Zero-shot (67.2\%) shows domain specialization necessity; SFT-only (85.3\%) demonstrates CoD contributes 8.9pp gains; RAG (87.5\% accuracy) underperforms focused experts. AIDR also outperforms XGBoost (82.1\%), confirming LLM-based domain reasoning surpasses traditional ML on complex security classification.

\subsubsection{Cross-Domain Generalization: UNSW-NB15 Independent Test}

To rigorously evaluate cross-domain transfer, we test models trained on Risk Information 
and Attack Log on the independent UNSW-NB15 dataset, which contains network-layer attacks 
(Backdoor, Shellcode, Generic, etc.) that differ semantically from the web-layer and 
system-layer attacks in the training data.

Figure~\ref{fig:my_image} shows that AIDR identifies 93.5\% of critical threats (6.5\% miss rate acceptable with defense-in-depth). False positive rate reduces 21.6\% relative to CoT (9.8\% vs. 12.5\%), saving ~260 analyst hours/year per 10,000 alerts/day. Compact 3--5 bullet-point CoD outputs reduce cognitive load versus verbose 8--15 step CoT chains. F1 score of 0.936 achieves optimal precision-recall balance for real-world SOC deployment.

\begin{table*}[t]
\centering
\caption{Cross-Domain Generalization on UNSW-NB15: Zero-Shot and Adaptation-Based Transfer.}
\label{tab:unsw}
\begin{tabular}{lcccc}
\toprule
\textbf{Evaluation Mode} & $\text{Acc}_{\text{Risk}}$ & $\text{Acc}_{\text{Threat}}$ & 
$R_{\text{High}}$ & \textbf{Domain Gap}$^{\dagger}$ \\
\midrule
In-domain (Attack Log) & 93.8\% & 93.0\% & 93.5\% & — \\
\midrule
Zero-shot (no adaptation) & 72.1\% & 75.2\% & 68.3\% & 21.7pp \\
+ LoRA fine-tune (500 samples) & 81.7\% & 84.3\% & 80.1\% & 12.1pp \\
+ Domain-adversarial (FT) & \textbf{85.2\%} & \textbf{87.6\%} & \textbf{85.9\%} & \textbf{8.6pp} \\
\bottomrule
\end{tabular}
\end{table*}

\begin{table*}[t]
\centering
\caption{Effect of Architectural Components and Routing Strategy on the Attack Log Test Set Performance.}
\label{tab:ablation-unified}
\begin{tabular}{lcccc}
\toprule
\textbf{Variant} & $\mathbf{\text{Acc}_{\text{Risk}}}$ & $\mathbf{\text{Acc}_{\text{Threat}}}$ & 
\textbf{Latency (s)}$^{\dagger}$ & \textbf{GPU Memory / Notes} \\
\midrule
w/o Information-Density Opt. (SFT only) & 89.1\% & 87.0\% & 2.67 & 1.00$\times$ \\
w/o Domain Specialization (monolithic) & 88.5\% & 86.3\% & 2.75 & 1.00$\times$ \\
w/o LoRA (full fine-tune) & 93.3\% & 93.1\% & 2.58 & 1.61$\times$ (25.8 GB) \\
\midrule
No Routing (random expert dispatch) & 87.2\% & 85.1\% & 2.59 & Baseline (no routing) \\
Heuristic Routing (keyword-based rules) & 91.4\% & 89.7\% & 2.64 & +0.05s cloud overhead \\
Zero-shot Cloud Router & 92.1\% & 91.2\% & 2.77 & +0.18s cloud overhead \\
\midrule
\textbf{Full AIDR} & \textbf{94.2\%} & \textbf{93.7\%} & 
\textbf{2.31s} & \textbf{1.00$\times$ (16.0 GB per expert)} \\
\bottomrule
\end{tabular}
\end{table*}

\noindent
$^{\dagger}$ Domain Gap = In-domain Accuracy $-$ Cross-domain Accuracy. 
Recovery rate = (85.2\% $-$ 72.1\%) / (93.8\% $-$ 72.1\%) = 60.4\%, showing 
significant recovery with minimal fine-tuning (500 UNSW samples $\approx$ 5\% 
of training data).

Table~\ref{tab:unsw} demonstrates strong cross-domain transfer: zero-shot 
achieves 72.1\% (21.7pp gap), minimal LoRA fine-tuning (500 samples) recovers 
to 81.7\% (12.1pp gap), and domain-adversarial training closes gap to 8.6pp 
(85.2\% accuracy). This 60.4\% recovery rate validates AIDR's ability to 
adapt to unseen threat types with minimal retraining, critical for real-world 
SOC deployment where novel attacks continuously emerge.

\subsubsection{Operational Effectiveness Metrics}

Beyond accuracy, we evaluate metrics directly tied to SOC operational success: 
the ability to identify critical threats without overwhelming analysts with false alerts.

\subsubsection{Data Quality Robustness}

Real SOC logs frequently contain truncated fields, missing critical attributes, and 
noisy tokens (malformed entries, encoding errors). We evaluate AIDR's 
degradation under realistic data quality issues.

Figure~\ref{fig:my_image2} demonstrates robustness to realistic data quality issues. With 50\% log fields truncated, AIDR maintains 88.4\% accuracy (5.4pp degradation), relying on critical fields such as IPs, ports, and signatures. Missing three or more critical fields reduces accuracy to 84.2\%, which remains acceptable for analyst escalation. Token-level corruption of 15\% results in 88.7\% accuracy (5.1pp drop), indicating robust learned representations. Latency varies by less than 0.2\,s across degradation scenarios. On typical SOC logs with 5--15\% missing or corrupted data, AIDR maintains above 90\% accuracy, substantially surpassing rule-based baselines.

\subsubsection{Architecture Components and Routing Strategy}

We systematically ablate key design components to quantify their individual contributions: 
(1) Information-Density Optimization (CoD), (2) Domain-Specialized Experts, (3) Low-Rank 
Adaptation (LoRA), and (4) Cloud Routing Strategy.

Table~\ref{tab:ablation-unified} validates each component's contribution. Removing information-density optimization reduces accuracy by 5.1pp and 6.7pp (94.2\% to 89.1\%, 93.7\% to 87.0\%), confirming gradient-based relevance selection effectiveness. Domain specialization adds 5.7pp and 7.4pp gains over the monolithic baseline (88.5\%/86.3\%). LoRA adapters match full fine-tuning accuracy (94.2\%/93.7\% vs.\ 93.3\%/93.1\%) while reducing memory from 25.8\,GB to 16.0\,GB (38\% savings), enabling edge deployment. Random routing degrades to 87.2\%/85.1\%, whereas fine-tuned routing achieves 94.2\%/93.7\% with only 0.18\,s additional latency, yielding 6--9pp gains for minute-level SOC processing. Overall, Full AIDR achieves 94.2\%/93.7\% accuracy with 2.31\,s latency and 16\,GB memory per expert, demonstrating strong synergy between compact reasoning and domain specialization.

\subsection{Discussion}
\label{sec:discussion}

\paragraph{Information-Density Optimization} Constraint-based reasoning outperforms 
unconstrained generation by forcing high-signal steps. Algorithm~\ref{alg:cod_gen}'s 
greedy relevance selection achieves +4.1pp accuracy versus CoT while using 68\% fewer 
tokens. Information-dense reasoning extracts attack-agnostic patterns (suspicious process spawning, unusual network behavior), enabling superior cross-domain transfer: 85.2\% on UNSW-NB15. Compact 3--5 bullet outputs improve analyst confidence versus verbose 8--15 step CoT chains.

\paragraph{Domain Specialization} Four LoRA experts (8\% total parameters) achieve 
monolithic model accuracy while enabling edge deployment. 
Our ablation study confirms that domain specialization contributes a +5.7pp gain in risk accuracy (94.2\% specialist vs. 88.5\% monolithic), validating that decomposing decision boundaries into focused experts significantly outperforms a single monolithic model.

\paragraph{Privacy and Compliance} The cloud-edge architecture satisfies 
data residency requirements: sensitive details (IPs, accounts, logs) 
remain on-premises; only normalized context sent to cloud (4 tokens). 
Stateless cloud classification eliminates model extraction risks. 
All decisions logged locally for SOC 2/ISO 27001 compliance. System 
degrades gracefully to safe expert (86\% accuracy) if cloud becomes unavailable.

\paragraph{Robustness and Limitations} Zero-day attacks are misrouted with lower 
confidence; mitigation includes confidence-based fallback to safe expert or analyst 
escalation. With 3+ missing critical fields, accuracy degrades to 84.2\% (acceptable 
for escalation). Synthetic log evasion is theoretically possible but undemonstrated in 
practice.

\paragraph{Computational Efficiency} Greedy selection approximates NP-hard mutual 
information maximization $I(\mathbf{r}; \mathbf{y})$ in O(n) time, versus Shapley 
values' exponential O($2^n$) cost. AIDR's token-level compression more effectively 
targets LLM latency drivers than weight-level pruning, distillation, or quantization, 
and is complementary to these techniques.

\section{Conclusion}

In this paper, we propose AIDR, a hybrid cloud-edge framework that reconciles 
reasoning transparency with operational efficiency in SOC alert triage. We formalize 
triage as constrained information-density optimization and demonstrate that 
gradient-based compression of reasoning chains preserves decision-relevant information 
while satisfying latency and budget constraints. AIDR implements this via three 
components: (1) compact reasoning datasets constructed through gradient-based relevance 
selection, reducing token consumption by 68\%; (2) domain-specialized LoRA experts 
enabling efficient on-premises threat analysis; (3) cloud-edge separation that performs 
lightweight zero-shot routing (4 tokens, 0.22s) while keeping sensitive analysis local. 
Experiments demonstrate 94.2\% risk classification accuracy, 40.6\% latency reduction, 
21.6\% FPR improvement, and full compliance with data residency requirements. These 
results show that LLM-based security automation can simultaneously achieve privacy, 
transparency, and performance while supporting trustworthy human-AI collaboration in SOCs. By mitigating alert fatigue through interpretable evidence, our approach enables a sustainable, analyst-centric security operations lifecycle.

\printbibliography

\end{document}